\documentclass[10pt,letterpaper]{article}
\usepackage{opex3}

 \usepackage{epstopdf}
 \usepackage{graphics}

\begin{document}

\title{Quantum entanglement distribution with 810 nm photons through active telecommunication fibers}
\author{Catherine Holloway, Evan Meyer-Scott, Chris Erven, Thomas Jennewein}
\address{Institute for Quantum Computing
University of Waterloo
200 University Ave. West
Waterloo, Ontario, CA
N2L 3G1 }

\email{c2hollow@iqc.ca}

\begin{abstract}
We demonstrate the distribution of polarization-entangled photons for the purpose of quantum key distribution (QKD) along active telecom fibers. Entangled photon pairs of 810~nm wavelength generated by a Sagnac interferometer source were coupled into standard telecom single mode fibers. The fibers were either dark or carrying a standardized 1550~nm ethernet signals (1000BASE-ZX) with a nominal speed of 1~GBps from regular media converter devices, without any  requirements on the optical power or spectrum transmitted.  Our system demonstrates a QKD network covering 6~km in distance with a central service provider for classical and quantum data. 
\end{abstract}



\bibliographystyle{osajnl}


\section*{Introduction}

Quantum key distribution (QKD) offers an alternative to existing key exchange protocols such as RSA, discrete logarithms\cite{paterson04} or trusted couriers  by using individual quanta as the information carriers, which cannot be altered without introducing errors to their state. QKD systems may either use discrete variables represented by single photons  (BB84 protocol\cite{bb84}), or entangled photons (BBM92 protocol\cite{PhysRevLett.68.557}) , or continuous variables represented in the quadratures of light\cite{Ralph:1999lr}. QKD has been demonstrated over free space links \cite{Erven:08, ursin2007entanglement, kurtsiefer2002quantum} as well as fiber links \cite{Poppe:04,Hubel:07}. 
\\
If QKD is to be adopted in real-world applications, it must be compatible with existing communications infrastructure. Most fiber-based implementations of QKD use `dark fibers' dedicated solely to quantum information \cite{hiskett2006long,stucki2002quantum}, an expensive usage of resources, or use parts of the optical spectrum which currently have low volume of traffic\cite{chapuran2009optical,1367-2630-12-6-063027,1367-2630-11-4-045012,lanchoqkd,choi2011}. Telecom wavelengths are used for the QKD signal in order to minimize attenuation, but lead to extra dark counts due to four-wave mixing processes between classical channels \cite{chapuran2009optical}. Noise cancellation or power regulation of the classical signal are required to counteract this effect. Additionally, difficult to operate InGaAs photon detectors or superconducting detectors must be used for these longer wavelength photons\cite{0957-0233-21-1-012002}.
\\
A multi-user network QKD system \cite{Lim:08}, such as a group of buildings within the same organization, with a central service provider for both classical and quantum data connections is envisioned. An entangled photon source is installed at this central location where the intranet connections between buildings are controlled in-house and are physically realized with  fiber-optic links at telecom wavelengths. When two entities want to make a secure connection, they request that the server room connect them to the source through the fiber optic links. Next, they each perform measurements on their entangled photons received from the source, and thus perform key growth according to the BBM92 protocol. 
\\
In this work, we show a simple and efficient solution for the compatibility of quantum and classical traffic without contraints on the classical traffic. We use polarization-entangled 810~nm photons travelling in 1550~nm single mode fibers (SMF). While it has been shown that entanglement-based
\cite{meyer2010quantum} and single photon QKD \cite{collins:073102}
are possible on dark telecom fibers; we now show that
entanglement-based QKD at visible wavelengths can readily be incorporated into existing active telecom fibers with a simple system. 810~nm photons are much more attenuated in optical fiber than 1550~nm (3 dB/km compared to 0.2 dB/km) which limits the total distance of the system, however at short distances we expect this wavelength to outperform infrared secure key rates due the use of higher efficiency SiAPD detectors.  
\section*{Implementation}
\begin{figure}
\includegraphics[width=15cm]{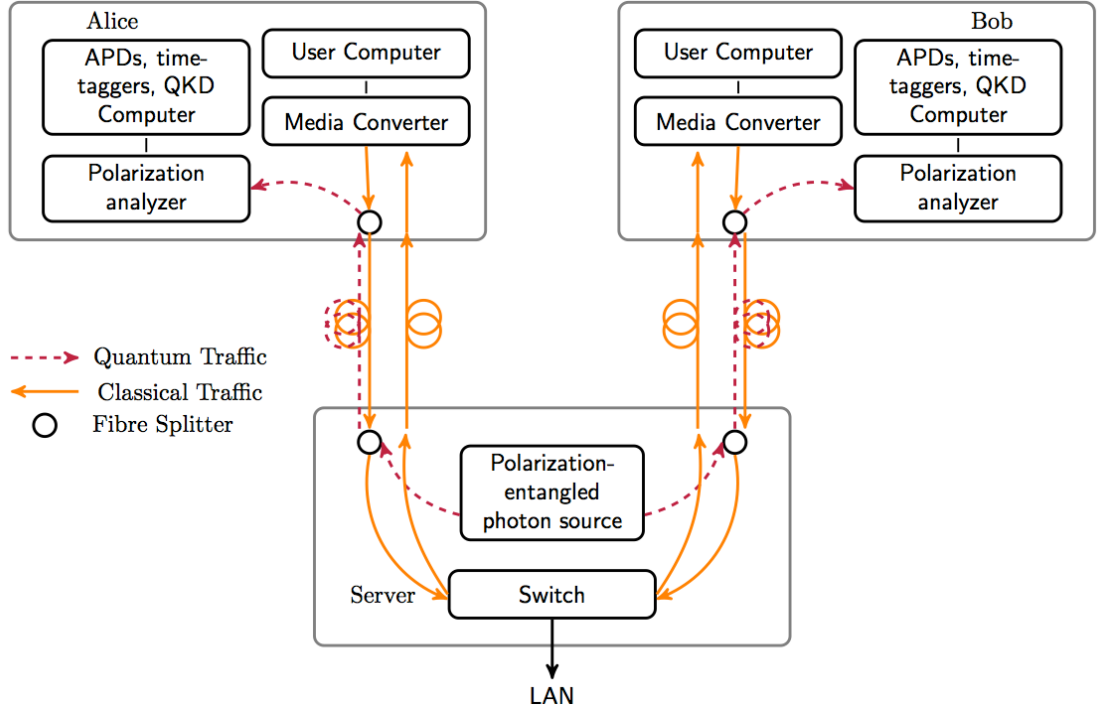}
\caption{Schematic of experimental apparatus. Two fiber splitters (represented by circles in this diagram) connect the 1550 nm fiber links. Classical information (represented by solid lines) travels in the opposite direction along the fiber as the quantum information (represented by dashed lines) and is generated by copper to optical media converters. For details on the polarization analyzer, see \cite{Erven:08}, for details on the photon source, see \cite{Fedrizzi:07}.}
\label{fig:experiment}
\end{figure}
As in Figure \ref{fig:experiment}, the two photons output by a Sagnac source \cite{Fedrizzi:07} were coupled into SMF for 810 nm and then into symmetric telecom fiber links and then sent to Alice and Bob for detection.  Fused biconical taper fiber beamsplitters in SMF were used to combine the gigabit ethernet signals at 1550~nm (from a 1000BASE-ZX transceiver with an optical power of 0.55 mW) with the 810~nm entangled photons. Commercially available common fiber splitters split 1550~nm signals equally among outputs, but the 810~nm photons couple only weakly to the branching port  and are mostly preserved \cite{photonics07}. The classical signals travelled along the fiber in the directions opposite to the quantum signals to minimize background counts in the quantum receivers. Traffic in the link was generated by using the secure copy protocol \cite{ssh} to transfer files on a local area network and measured using shallow packet inspection \cite{shallow}. 
\\
At both outputs of the fiber channels, 1550 nm SMF fiber splitters were used to extract the classical signals. The other outputs of the fiber splitters were passed through a 5 nm wide bandpass \@ 810~nm filter to remove stray telecom signal and then through a polarization analyzer which split the photons to four different silicon photon avalanche photodiodes (Si-APD, Perkin-Elmer SPCM Quad Array) based on their polarization (rectilinear (H,V) or diagonal (+,-)). The APDs were attached to a timetagging unit which passed the time and state of the signals to Alice and Bob's computers for post-processing. 
\\
Waveguide theory predicts that two orders for the spatial modes of 810~nm light may propagate in 1550~nm SMF \cite{photonics07}, and these will differ in group velocity by about 2.2~ns per km of fiber \cite{681313}. Spatial and temporal filtering was performed to exclude the second-order propagation mode from the first-order mode, by using the mode selectivity of a  2~m single-mode fiber for 810 nm applied just before detection of the photons, and post-processing the arrival times of the photons to filter out the photons arriving in the delayed second-order mode \cite{meyer2010quantum}. This successfully increased the received entanglement visibility from 62\% (unfiltered) to 95\% (filtered) at the cost of only about half of photon pairs.  
\\
Alice and Bob communicate along a separate classical link to find the time offset needed to maximize the coincidences in their detection events. Thus they can compare detection events to calculate the entanglement visibility
and the quantum bit error rate (QBER). The QBER is an important measure because it ensures the integrity of the secrecy of the communication \cite{RevModPhys.74.145} - to maintain the security of the key distribution, the QBER should be at least less than 11\%, the threshold for QKD in the infinite key limit\cite{scarani-2009-81}.
\\
Coincident detection rates and QBERs were measured for total fiber lengths between 0.5 and 6 km for sufficient duration that the minimum number of raw key bits required to generate a secret key in the finite key limit\cite{finitesize} were obtained.
\section*{Results}
\begin{figure}
\includegraphics[width=16cm]{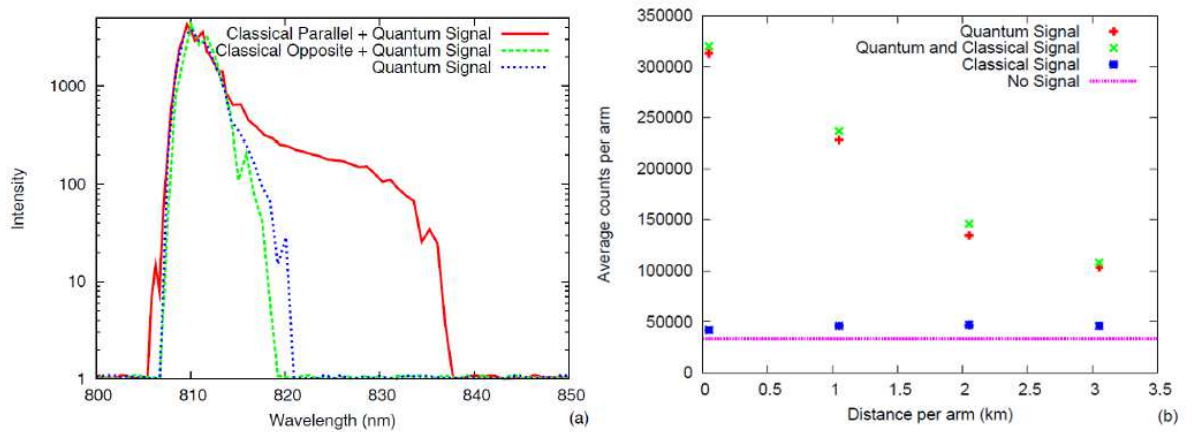}
\caption{Signal spectrum in (a) is intensity averaged in bins of 1 nm, for various signal configurations on 2 km of 1550~nm SMF. (b) is the total detected counts per second with classical signal opposite. In (b), the counts in the no-signal case are the total detector dark counts per arm. Statistical error on the count number is smaller than the size of the data point.}
\label{fig:spectra}
\end{figure}
\subsection*{System Characterization}
We observed the light after fiber transmission using a spectrometer in order to quantify the interaction of our short-wavelength quantum and long-wavelength classical signals. Measurements of the spectrum of the quantum signal were taken with no classical signal, and with the classical signal in either parallel or opposite propagation directions on a 2 km long fiber. Results of these measurements are presented in figure \ref{fig:spectra}.
\\
The parallel propagation direction introduced more photons into the fiber around the wavelength of the quantum signal, between 820 and 840 nm. This indicates that some mixing process occured between the quantum signal at 810 nm and the classical signal at 1550 nm. The opposite propagation direction introduced negligible background photons to our quantum signal, and so this propagation direction was used in experiment. The classical signal produced on average 500 extra counts per detector, regardless of the length of the fiber or the presence of quantum signal. For implementations using optical communication standards other than what we used, further investigation of the background suppression is required.
\\
From the coincidence and single detection rates, we estimate that our source is producing entangled pairs at a rate of 0.4 MHz.
\subsection*{QKD Results}
\begin{figure}
\includegraphics[width=16cm]{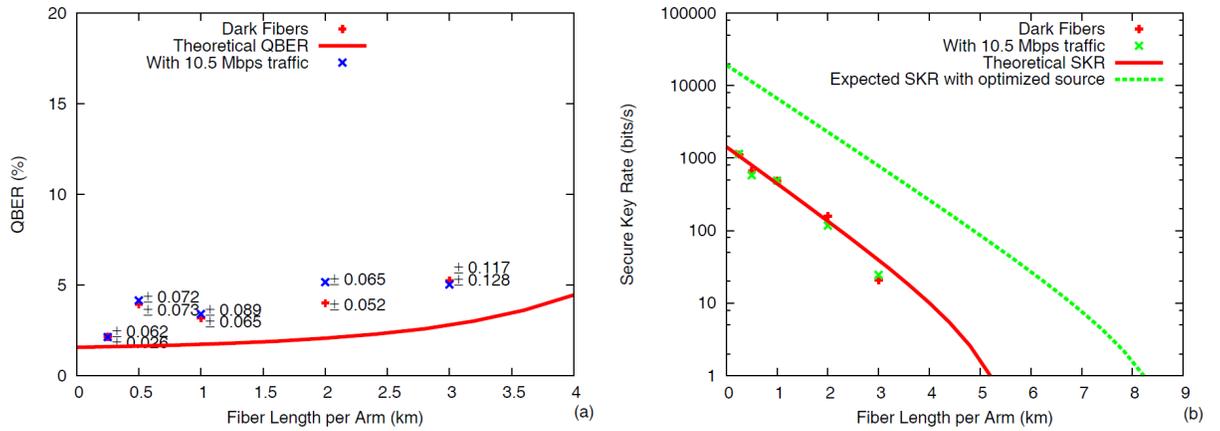}
\caption{QBER (a) and SKR (b) for varying lengths of fiber per arm, which are dark (no telecom signals introduced) or with telecom signals carrying 10.5 Mbps of traffic. The observed QBER is larger than predicted by theory at longer lengths due to the difficulty in aligning polarization bases with attenuated count rates, even at the maximum source rate. SKR was simulated using the mathematical descriptions of detectors and SPDC sources of entangled photons described in \cite{PhysRevA.76.012307}, with experimental and ideal parameters.}
\label{fig:qber}
\end{figure}

\begin{figure}

\includegraphics[width=8cm]{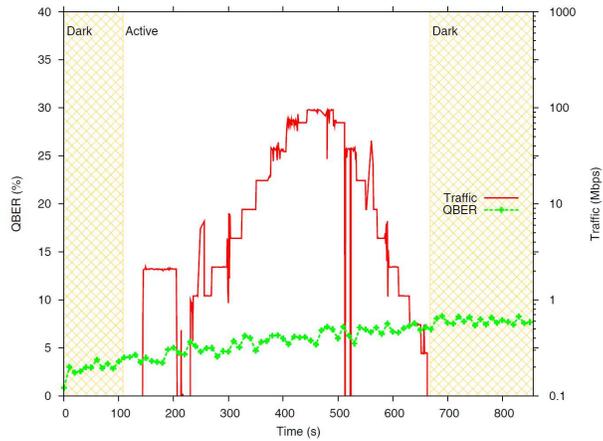}
\caption{QBER for varying levels of telecom traffic, averaged in bins of 10 s. The QBER increases slightly over time due to polarization drift, but is unaffected by classical traffic. }
\label{fig:traffic}
\end{figure}

The observed QBER versus fiber length between the server and Alice or Bob for both active and dark fibers is presented in figure \ref{fig:qber}. At all lengths measured (0.25, 0.5, 1, 2, and 3 km for each channel) the QBER for active fibers was the same to within a few percentage points as the QBER for dark fibers. 
\\
All active-fiber experiments were conducted with around 10 Mbps of traffic on the fibers. An additional experiment was conducted at 4~km each between the server and Alice or Bob, where the traffic was varied between 0 and 100 Mbps, with negligible effect on the QBER (figure \ref{fig:traffic}). The  media converters used in this experiment transmitted with the same intensity on idle as at 100 Mbps so the QBER was expected to be independent of traffic. 
\\
From the measured QBER we  calculated the lower bound on the estimated secure key rate \cite{PhysRevA.76.012307}. Simulations of secure key rate for the fiber lengths used in this experiment with appropriate variables were calculated \cite{jenneweinjmo:11} and compared to secure key rates estimated with experimentally obtained QBER. Simulated QBER was larger than measured due to relaxation in polarization alignment in the fibers. QBER increased significantly at 3~km fiber lengths due the the difficulty in aligning polarization with few photon detector counts.  Asymptotic analysis \cite{PhysRevA.76.012307} predicts positive secure key rate up to 6km fiber lengths but the increased QBER made it impossible to get secure keys with finite size statistics \cite{finitesize}.
\section*{Conclusion}
We have demonstrated high-visibility key distribution over active telecom fibers over total fiber lengths up to 3~km, and with symmetric links this leads to total distances of 6~km, without disruption of either classical or quantum signal. The standard we have used is not as common for short links, which operate at 1310~nm, but a brief investigation of our system on an installed 2~km link  operating at 1310~nm did not result in any difficulties. Further investigation of the mixing process between quantum and classical signal may be required for optical communication standards where the spectrum is different than the one we used. We expect that this distance can be extended in the future, since simulations indicate that if the entangled source could operate at 15~Mpairs/s, which is about one order of magnitude greater then  our current Sagnac source but within the range of pair rates generated by previous experiments \cite{ursin2007entanglement}, the quantum signals could span a distance of 16~km. 
\\
We believe the system demonstrated here is a very elegant and effective approach for creating a multi-node QKD network. Integration of QKD into existing infrastructures means that security-critical organizations spanning many kilometers of diameter, such as within a city center or a campus, could ensure that their information is indefinitely secure. Note that our combination of the QKD signals with the classical signal can be achieved with very simple, robust and cost-effective technology, and does not place restrictions on the classical traffic. QKD systems using our approach can be easily incoporated into existing active single mode fiber networks, only requiring minor alterations to existing communication channels. 

\section*{Acknowledgements}
The authors acknowledge useful conversations with Colin Bell, Alessandro Fedrizzi, Brendon Higgins, and Hannes H\"{u}bel, and funding from NSERC (CGS, QuantumWorks, Discovery, USRA), Ontario Ministry of Research and Innovation (ERA program), CIFAR, Industry Canada, the Canadian Space Agency and the CFI.


\begin{thebibliography}{10}
\newcommand{\enquote}[1]{``#1''}


\bibitem{paterson04}
K.~G. Paterson, F.~{Piper}, and R.~{Schack}, \enquote{{Quantum cryptography: a
  practical information security perspective},} ArXiv Quantum Physics e-prints
  (2004).

\bibitem{bb84}
C.~H. Bennett and G.~Brassard, \enquote{Quantum cryptography: Public key
  distribution and coin tossing,} in \enquote{Proceedings of the IEEE
  International Conference on Computers, Systems, and Signal Processing,}
  (1984), p. 175.

\bibitem{PhysRevLett.68.557}
C.~H. Bennett, G.~Brassard, and N.~D. Mermin, \enquote{Quantum cryptography
  without bell's theorem,} Phys. Rev. Lett. \textbf{68}, 557--559 (1992).

\bibitem{Ralph:1999lr}
T.~C. Ralph, \enquote{Continuous variable quantum cryptography,} Phys. Rev. A
  \textbf{61}, 010303 (1999).

\bibitem{Erven:08}
C.~Erven, C.~Couteau, R.~Laflamme, and G.~Weihs, \enquote{Entangled quantum key
  distributionover two free-space optical links,} Opt. Express \textbf{16},
  16840--16853 (2008).

\bibitem{ursin2007entanglement}
R.~Ursin, F.~Tiefenbacher, T.~Schmitt-Manderbach, H.~Weier, T.~Scheidl,
  M.~Lindenthal, B.~Blauensteiner, T.~Jennewein, J.~Perdigues, P.~Trojek
  \emph{et~al.}, \enquote{Entanglement-based quantum communication over 144
  km,} Nature Physics \textbf{3}, 481--486 (2007).

\bibitem{kurtsiefer2002quantum}
C.~Kurtsiefer, P.~Zarda, M.~Halder, H.~Weinfurter, P.~Gorman, P.~Tapster, and
  J.~Rarity, \enquote{Quantum cryptography: A step towards global key
  distribution,} Nature \textbf{419}, 450--450 (2002).

\bibitem{Poppe:04}
A.~Poppe, A.~Fedrizzi, R.~Ursin, H.~B\"{o}hm, T.~L\"{o}runser, O.~Maurhardt,
  M.~Peev, M.~Suda, C.~Kurtsiefer, H.~Weinfurter, T.~Jennewein, and
  A.~Zeilinger, \enquote{Practical quantum key distribution with polarization
  entangled photons,} Opt. Express \textbf{12}, 3865--3871 (2004).

\bibitem{Hubel:07}
H.~H\"{u}bel, M.~R. Vanner, T.~Lederer, B.~Blauensteiner, T.~Lor\"{u}nser,
  A.~Poppe, and A.~Zeilinger, \enquote{High-fidelity transmission of
  polarization encoded qubits from an entangled source over 100 km of fiber,}
  Opt. Express \textbf{15}, 7853--7862 (2007).

\bibitem{hiskett2006long}
P.~Hiskett, D.~Rosenberg, C.~Peterson, R.~Hughes, S.~Nam, A.~Lita, A.~Miller,
  and J.~Nordholt, \enquote{Long-distance quantum key distribution in optical
  fibre,} New Journal of Physics \textbf{8}, 193 (2006).

\bibitem{stucki2002quantum}
D.~Stucki, N.~Gisin, O.~Guinnard, G.~Ribordy, and H.~Zbinden, \enquote{Quantum
  key distribution over 67 km with a plug\&play system,} New Journal of Physics
  \textbf{4}, 41 (2002).

\bibitem{chapuran2009optical}
T.~Chapuran, P.~Toliver, N.~Peters, J.~Jackel, M.~Goodman, R.~Runser,
  S.~McNown, N.~Dallmann, R.~Hughes, K.~McCabe \emph{et~al.}, \enquote{Optical
  networking for quantum key distribution and quantum communications,} New
  Journal of Physics \textbf{11}, 105001 (2009).

\bibitem{1367-2630-12-6-063027}
P.~Eraerds, N.~Walenta, M.~Legr{\'e}, N.~Gisin, and H.~Zbinden,
  \enquote{Quantum key distribution and 1 gbps data encryption over a single
  fibre,} New Journal of Physics \textbf{12}, 063027 (2010).

\bibitem{1367-2630-11-4-045012}
N.~A. Peters, P.~Toliver, T.~E. Chapuran, R.~J. Runser, S.~R. McNown, C.~G.
  Peterson, D.~Rosenberg, N.~Dallmann, R.~J. Hughes, K.~P. McCabe, J.~E.
  Nordholt, and K.~T. Tyagi, \enquote{Dense wavelength multiplexing of 1550 nm
  qkd with strong classical channels in reconfigurable networking
  environments,} New Journal of Physics \textbf{11}, 045012 (2009).

\bibitem{lanchoqkd}
D.~Lancho, J.~Martinez, D.~Elkouss, M.~Soto, and V.~Martin, \enquote{Qkd in
  standard optical telecommunications networks,} in \enquote{Quantum
  Communication and Quantum Networking,} , vol.~36 of \emph{Lecture Notes of
  the Institute for Computer Sciences, Social Informatics and
  Telecommunications Engineering}, O.~Akan, P.~Bellavista, J.~Cao, F.~Dressler,
  D.~Ferrari, M.~Gerla, H.~Kobayashi, S.~Palazzo, S.~Sahni, X.~S. Shen,
  M.~Stan, J.~Xiaohua, A.~Zomaya, G.~Coulson, A.~Sergienko, S.~Pascazio, and
  P.~Villoresi, eds. (Springer Berlin Heidelberg, 2010), pp. 142--149.

\bibitem{choi2011}
I. ~Choi, R. ~J. Young, and P. ~D. Townsend, \enquote{Quantum information to the home,} 
New Journal of Physics \textbf{13}, 063039 (2011).

\bibitem{0957-0233-21-1-012002}
G.~S. Buller and R.~J. Collins, \enquote{Single-photon generation and
  detection,} Measurement Science and Technology \textbf{21}, 012002 (2010).

\bibitem{Lim:08}
H.~C. Lim, A.~Yoshizawa, H.~Tsuchida, and K.~Kikuchi, \enquote{Distribution of
  polarization-entangled photonpairsproduced via spontaneous
  parametricdown-conversion within a local-area fibernetwork: Theoretical model
  and experiment,} Opt. Express \textbf{16}, 14512--14523 (2008).

\bibitem{meyer2010quantum}
E.~Meyer-Scott, H.~Hubel, A.~Fedrizzi, C.~Erven, G.~Weihs, and T.~Jennewein,
  \enquote{Quantum entanglement distribution with 810 nm photons through
  telecom fibers,} Applied Physics Letters \textbf{97}, 031117--031117 (2010).

\bibitem{collins:073102}
R.~J. Collins, P.~J. Clarke, V.~Fern\'{a}ndez, K.~J. Gordon, M.~N. Makhonin,
  J.~A. Timpson, A.~Tahraoui, M.~Hopkinson, A.~M. Fox, M.~S. Skolnick, and
  G.~S. Buller, \enquote{Quantum key distribution system in standard
  telecommunications fiber using a short wavelength single photon source,}
  Journal of Applied Physics \textbf{107}, 073102 (2010).

\bibitem{Fedrizzi:07}
A.~Fedrizzi, T.~Herbst, A.~Poppe, T.~Jennewein, and A.~Zeilinger, \enquote{A
  wavelength-tunable fiber-coupled source of narrowband entangled photons,}
  Opt. Express \textbf{15}, 15377--15386 (2007).

\bibitem{ssh}
D.~J. Barrett, R.~E. Silverman, R.~G. Byrnes,
\newblock {\sl SSH, the secure shell: the definitive guide}, 
\newblock (O'Reilly Media, Inc., 2005)

\bibitem{shallow}
J.~Matthews,
\newblock {\sl Computer Networking: Internet Protocols in Action}, 
\newblock (Wiley, 2005)


\bibitem{photonics07}
B.~E.~A. Saleh and M.~C. Teich, \emph{Fundamentals of Photonics} (Wiley, 2007),
  2nd ed.

\bibitem{681313}
P.~Townsend, \enquote{Experimental investigation of the performance limits for
  first telecommunications-window quantum cryptography systems,} Photonics
  Technology Letters, IEEE \textbf{10}, 1048 --1050 (1998).

\bibitem{RevModPhys.74.145}
N.~Gisin, G.~Ribordy, W.~Tittel, and H.~Zbinden, \enquote{Quantum
  cryptography,} Rev. Mod. Phys. \textbf{74}, 145--195 (2002).

\bibitem{scarani-2009-81}
V.~Scarani, H.~Bechmann-Pasquinucci, N.~J. Cerf, M.~Dusek, N.~Lutkenhaus, and
  M.~Peev, \enquote{The security of practical quantum key distribution,}
  Reviews of Modern Physics \textbf{81}, 1301 (2009).

\bibitem{finitesize}
V.~Scarani, and R.~Renato,\enquote{Quantum Cryptography with Finite Resources: Unconditional Security Bound for Discrete-Variable Protocols with One-Way Postprocessing}, Phys. Rev. Lett. \textbf{20}, 200501 (2008).

\bibitem{PhysRevA.76.012307}
X.~Ma, C.-H.~F. Fung, and H.-K. Lo, \enquote{Quantum key distribution with
  entangled photon sources,} Phys. Rev. A \textbf{76}, 012307 (2007).

\bibitem{jenneweinjmo:11}
T.~Jennewein, M.~Barbieri, and A.~G. White, \enquote{Single-photon device
  requirements for operating linear optics quantum computing outside the
  post-selection basis,} Journal of Modern Optics \textbf{58}, 276 -- 287
  (2011).

\end{thebibliography}
\end{document}